\documentclass[prl,twocolumn,amsfonts,superscriptaddress,showpacs]{revtex4}
\usepackage{graphicx}
\usepackage{color}
\usepackage{portland}

\begin{document}

\title{Evidence for a Peierls phase-transition in a three-dimensional multiple charge-density waves solid.}

\author{Barbara Mansart}
\affiliation{Laboratory for Ultrafast Electrons, ICMP, Ecole Polytechnique F\'{e}d\'{e}rale de Lausanne, CH-1015 Lausanne, CH}
\affiliation{Laboratory of Ultrafast Spectroscopy, ISIC, Ecole Polytechnique F\'{e}d\'{e}rale de Lausanne, CH-1015 Lausanne, CH}
\author{Mathieu J.G. Cottet}
\affiliation{Laboratory for Ultrafast Electrons, ICMP, Ecole Polytechnique F\'{e}d\'{e}rale de Lausanne, CH-1015 Lausanne, CH}
\author{Thomas J. Penfold}
\affiliation{Laboratory of Ultrafast Spectroscopy, ISIC, Ecole Polytechnique F\'{e}d\'{e}rale de Lausanne, CH-1015 Lausanne, CH}
\affiliation{Laboratory of Computational Chemistry and Biochemistry, ISIC, Ecole Polytechnique F\'{e}d\'{e}rale de Lausanne, CH-1015 Lausanne, CH}
\affiliation{SwissFEL, PSI, CH-5232 Villigen, CH}
\author{Stephen B. Dugdale}
\affiliation{H. H. Wills Physics Laboratory, University of Bristol, Tyndall Avenue, Bristol BS8 1TL, UK}
\author{Riccardo Tediosi}
\affiliation{D\'epartement de Physique de la Mati\`ere Condens\'ee, Universit\'e de Gen\`eve, CH-1211 Gen\`eve 4, CH}
\author{Majed Chergui}
\affiliation{Laboratory of Ultrafast Spectroscopy, ISIC, Ecole Polytechnique F\'{e}d\'{e}rale de Lausanne, CH-1015 Lausanne, CH}
\author{Fabrizio Carbone}
\affiliation{Laboratory for Ultrafast Electrons, ICMP, Ecole Polytechnique F\'{e}d\'{e}rale de Lausanne, CH-1015 Lausanne, CH}

\begin{abstract}

 The effect of dimensionality on materials properties has become strikingly evident with the recent discovery of graphene. Charge ordering phenomena can be induced in one dimension by periodic distortions of a material's crystal structure, termed Peierls ordering transition. Charge-density waves can also be induced in solids by strong Coulomb repulsion between carriers, and at the extreme limit, Wigner predicted that crystallization itself can be induced in an electrons gas in free space close to the absolute zero of temperature. Similar phenomena are observed also in higher dimensions, but the microscopic description of the corresponding phase transition is often controversial, and remains an open field of research for fundamental physics. Here, we photoinduce the melting of the charge ordering in a complex three-dimensional solid and monitor the consequent charge redistribution by probing the optical response over a broad spectral range with ultrashort laser pulses. Although the photoinduced electronic temperature far exceeds the critical value, the charge-density wave is preserved until the lattice is sufficiently distorted to induce the phase transition. Combining this result with {\it ab initio} electronic structure calculations, we identified the Peierls origin of multiple charge-density waves in a three-dimensional system for the first time.
 
\end{abstract}
\maketitle

\begin{widetext}

\large{Accepted for publication in Proc. Natl. Acad. Sci. USA, doi:10.1073/pnas.1117028109}
~\\
~\end{widetext}

Charge ordering phenomena occurring upon symmetry breaking are important in solids as they give rise to new current and spin flow patterns in promising materials such as organic conductors~\cite{Lorenz2002}, multilayered graphene~\cite{Ho2010} and transition metal oxides~\cite{Qazilbash2011}. The possibility to investigate the microscopic steps through which such ordering transition occurs gives also the opportunity to speculate on more general aspects of critical phenomena. Charge-density waves (CDWs)~\cite{Middleton1992, Narayan1994}, sandpile automata~\cite{tadic} and Josephson arrays~\cite{tinkham} have been investigated in relation to the scale invariance of self-organized critical phenomena~\cite{Bak1987}, of which avalanches are dramatic manifestations~\cite{vajont}.
 In one dimension, Peierls demonstrated that at low temperature an instability can be induced by the coupling between carriers and a periodic lattice distortion. Such an instability triggers a charge ordering phenomenon and a metal-insulator phase transition, called Peierls transition, occurs \cite{peierls}. Like for Bardeen-Cooper-Schrieffer (BCS) superconductors, such an electron-phonon interaction driven transition is expected to be second order \cite{peierls}. While this situation is fairly established in mono-dimensional organic materials \cite{tcnq}, increased hybridization leading to higher dimensionality of a solid perturbs this scenario and makes the assessment of the microscopic origin of charge localization phenomena more difficult~\cite{nbse3, tas2, mcmillan}.

 Contrary to other low-dimensional CDW~\cite{gruner} systems studied so far by time-resolved spectroscopies~\cite{yusupov, Eichberger2010,Schmitt2008, Tomeljak2009}, Lu$_5$Ir$_4$Si$_{10}$ presents a complex three-dimensional structure with several substructures such as one-dimensional Lu chains and three-dimensional cages in which a variety of many-body effects (including superconductivity) originate~\cite{smaalen} (Fig.~\ref{reflectivity}\textit{A}). Although a CDW occurs below $T_{CDW}$= 83 K, its microscopic origin is still debated since this transition is first order and isotropic without hysteresis~\cite{Becker1999, Tediosi}, in contrast to the standard Peierls paradigm. Also, owing to their complexity, a detailed microscopic description of the properties of these solids is still lacking and their band structure has never been reported.

The charge redistribution induced by a phase transition in a solid can be obtained via the optical frequency-sum (\textit{f-sum}) rule~\cite{Tediosi}, which states that the integral over all frequencies of the optical conductivity ($\sigma_1(\omega)$), termed spectral weight ($SW$), is constant and corresponds to the number of electrons per unit cell:

\begin{equation}
SW=\frac{\pi e^2}{2 m_e V} N_{eff}=\int^{\infty}_{0} \sigma_1 (\omega)d\omega
\label{Neff}
\end{equation}

where $m_e$ is the free electron mass, $V$ the unit cell volume and $N_{eff}$ the total number of carriers. Performing the integral up to a frequency cut-off $\Omega_c$ (partial sum-rule) gives the number of carriers contained in the electronic levels included in this energy range~\cite{SI}. Optical weight redistributions are known to accompany every BCS-like phase transition, as described by the Ferrel-Glover-Tinkham sum-rule \cite{fgt}.
Recently, it has been shown that information on the $SW$ can be obtained in a model-independent fashion based on the analytical continuation of holomorphic functions~\cite{devin}.

In this study we aim at obtaining the temporal evolution of the partial sum rule during the photoinduced annihilation of the CDW order parameter to identify the different states through which the electronic and the lattice structure evolve. To achieve this, the changes in the optical constants are monitored over a broad energy range with a temporal resolution better than the electron-phonon coupling time.
We performed pump-probe reflectivity using a broad (1.5-3 eV) supercontinuum of 50 fs polarized pulses as a probe~\cite{SI}. The temperature of the sample was controlled between 10 K and room temperature.  The 1.55 eV pump fluence was varied between 0.8 and 3.1 mJ/cm$^2$, corresponding to an absorbed fluence between 0.4 and 1.55 mJ/cm$^2$; in these conditions and for an initial temperature of 10 K, the electronic temperature reaches values between 302 and 1100 K (see the transient temperature analysis below), corresponding to energies in the range 26-94 meV. Therefore, for the lowest pump fluences, the absorbed energy lies below the CDW gap estimated by static optical spectroscopy (80 meV)~\cite{Tediosi}. Our ultrafast reflectivity data were combined with static optical spectroscopy~\cite{Tediosi} and new {\it ab initio} band structure calculations, allowing the identification of the spectroscopic signature of the CDW melting, the description of the orbitals and atomic motions involved and their temporal evolution.

\begin{widetext}

		\begin{figure}[ht]
		\vspace*{.05in}
\centerline{\includegraphics[width=150mm]{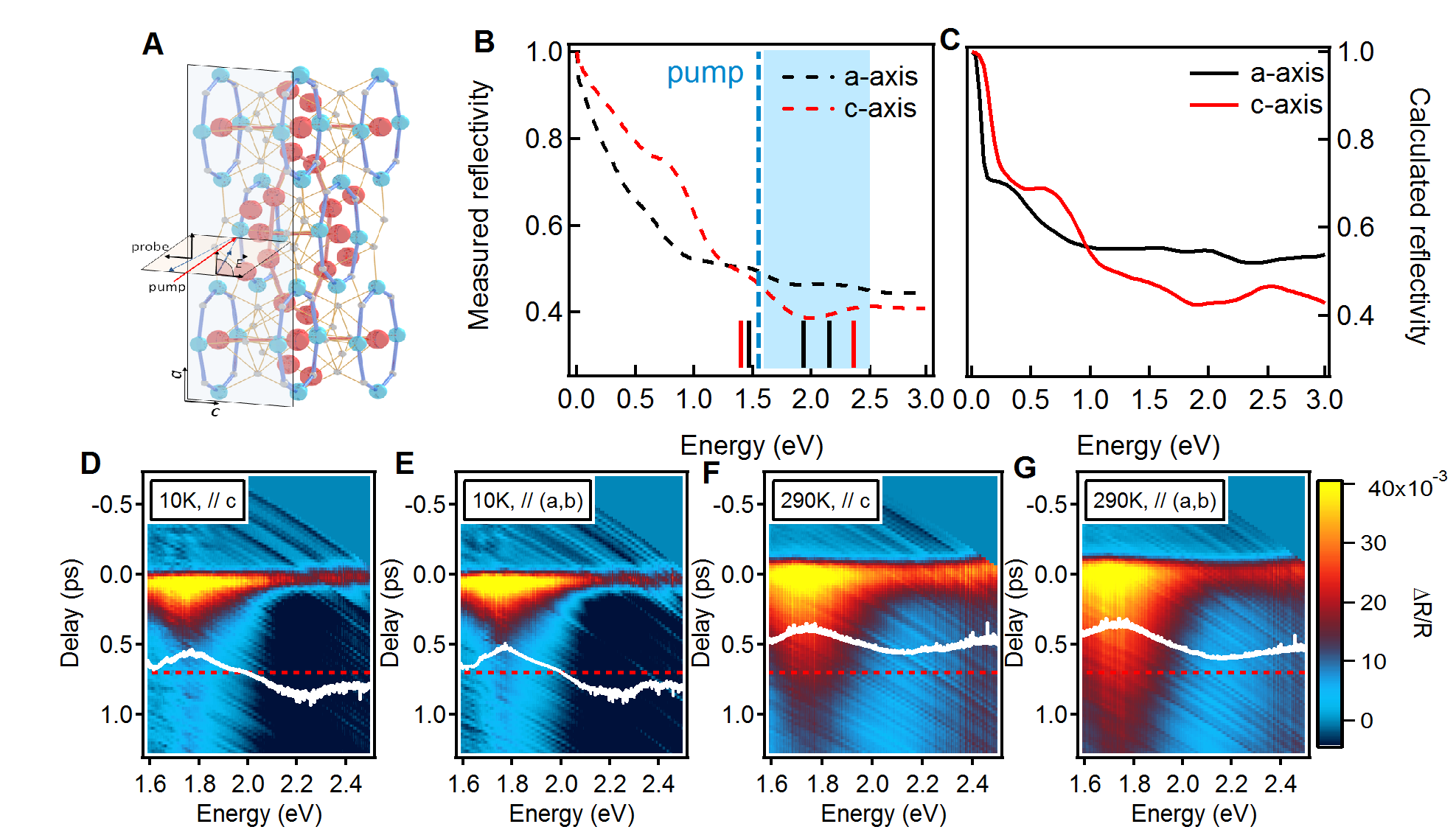}}
\caption{(\textit{A}) Experimental geometry and crystal structure; the pump and probe polarizations can be varied independently. (\textit{B}) Static reflectivity measurements; vertical bars indicate the optical absorptions of the solid, the vertical dashed line the energy of our photoexciting pulse, and the probe supercontinuum range is highlighted as a light-blue shaded area, covering a spectral region of large anisotropy. (\textit{C}) Calculated reflectivity along the $a$-axis and the $c$-axis. (\textit{D-G}) Time-resolved reflectivity images for a pump fluence of 3.1 mJ/cm$^2$ and (\textit{D-E}) T = 10 K, (\textit{F-G}) T = 290 K. White traces in (\textit{D-G}) are profiles extracted at 700 fs, whose baselines are shown as dashed red lines.}
\label{reflectivity}
\end{figure}

\end{widetext}

The static optical reflectivity of Lu$_{5}$Ir$_{4}$Si$_{10}$ for light polarized along the one-dimensional chains ($c$-axis), and perpendicular to them ($a$-axis) is displayed in Fig.~\ref{reflectivity}\textit{B}. Fig.~\ref{reflectivity}\textit{C} shows the calculated reflectivity which agrees with the experiments; some discrepancies are observed at low energy along the $c$-axis where correlation effects are expected due to the presence of the monodimensional chains.
Figs~\ref{reflectivity}\textit{(D-G)} show the transient optical reflectivity for different temperatures and light polarizations. The low temperature dynamic reflectivity decays faster than at room temperature (200 fs compared with 300 fs, for the same pump fluence and energy of 1.9 eV).
Its spectral behavior is also rather different: at 10 K, a large region is observed in which the photoinduced change in reflectivity switches sign (see white traces). The timescales and the absolute magnitude of the changes also depend on the polarization of the probe.


\begin{widetext}

\begin{figure}[ht]
\vspace*{.05in}
\centerline{\includegraphics[width=120mm]{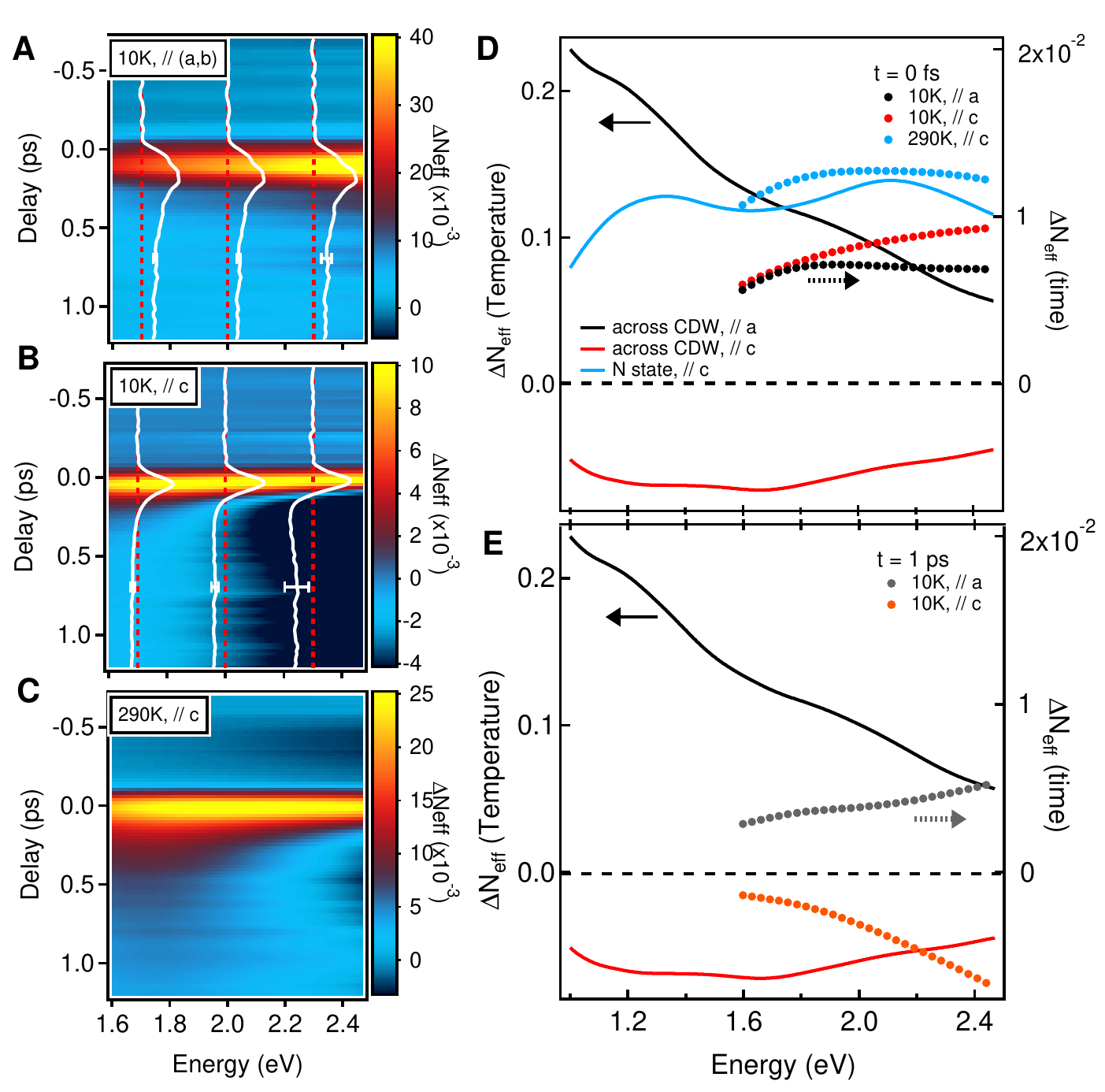}}
\caption{Spectral Weight analysis. (\textit{A-C}) 3D color maps of time-resolved $N_{eff}$. (\textit{D-E}) Difference in $N_{eff}$ between the normal ($T$ = 120 K) and the Charge-density wave  ($T$ = 60 K)  states for the two orientations, and between two different temperatures in the normal state (290 K - 120 K) (solid lines); time-resolved $N_{eff}$ for $T$ = 10 and 290 K, and probe parallel to $c$ or $a$-axis (circles), at time delays of 0 fs (\textit{D}) and 1 ps (\textit{E}).}
\label{SW}
\end{figure}

\end{widetext}

The evolution of $N_{eff}$ is obtained from the transient reflectivity both via a Drude-Lorentz model (3D color maps in Fig.~\ref{SW}) and the model-independent approach (white lines with error bars in Fig.~\ref{SW}\textit{A} and \textit{B})~\cite{SI}. In these plots, the integral in Eq.~\ref{Neff} is displayed as a function of the frequency cut-off, emphasizing the spectral regions of positive and negative change of the effective number of carriers. When the sample temperature is below $T_{CDW}$ and the probe light is polarized along the $c$-axis (Fig.~\ref{SW}\textit{B}), the $N_{eff}$ change, defined as ($N_{eff}$(t,$\omega$)-$N_{eff}$(t$<0,\omega$)), becomes negative after 200 fs. Instead, it is always positive when the polarization is perpendicular to the $c$-axis (Fig.~\ref{SW}\textit{A}), or at room temperature (Fig.~\ref{SW}\textit{C}). This sign change at low temperature along the $c$-axis signals a $SW$ transfer from energies higher than 1.5 eV, typical of strongly correlated solids~\cite{Rozenberg1996}. In Fig.~\ref{SW}\textit{D} and \textit{E}, the quantities ($N_{eff}$($T_2$,$\omega$)-$N_{eff}$($T_1$, $\omega$))/$N_{eff}$($T_1$,$\omega$) are displayed for $T_1$ = 60 K and $T_2$ = 120 K, with light polarized along the $c$-axis (red line) or the $a$-axis (black line), and $T_1$ = 120 K and $T_2$ = 290 K for light polarized along the $c$-axis (light blue line) (data taken from~\cite{Tediosi}). The $SW$ transfer across the CDW temperature along the $c$-axis becomes negative above 1.5 eV, as observed in the transient data; therefore, the spectroscopic signature of the CDW melting is the redistribution of carriers above 1.5 eV where bands composed of Lu and Ir $5d$ states are found. Remarkably, we observe that this feature appears several tens of fs after laser excitation, suggesting that before the CDW melts, an unconventional state of matter is produced in which some charge order survives despite the large energy deposited in the electronic subsystem.

\begin{widetext}

\begin{figure}[ht]
\vspace*{.05in}
\centerline{\includegraphics[width=120mm]{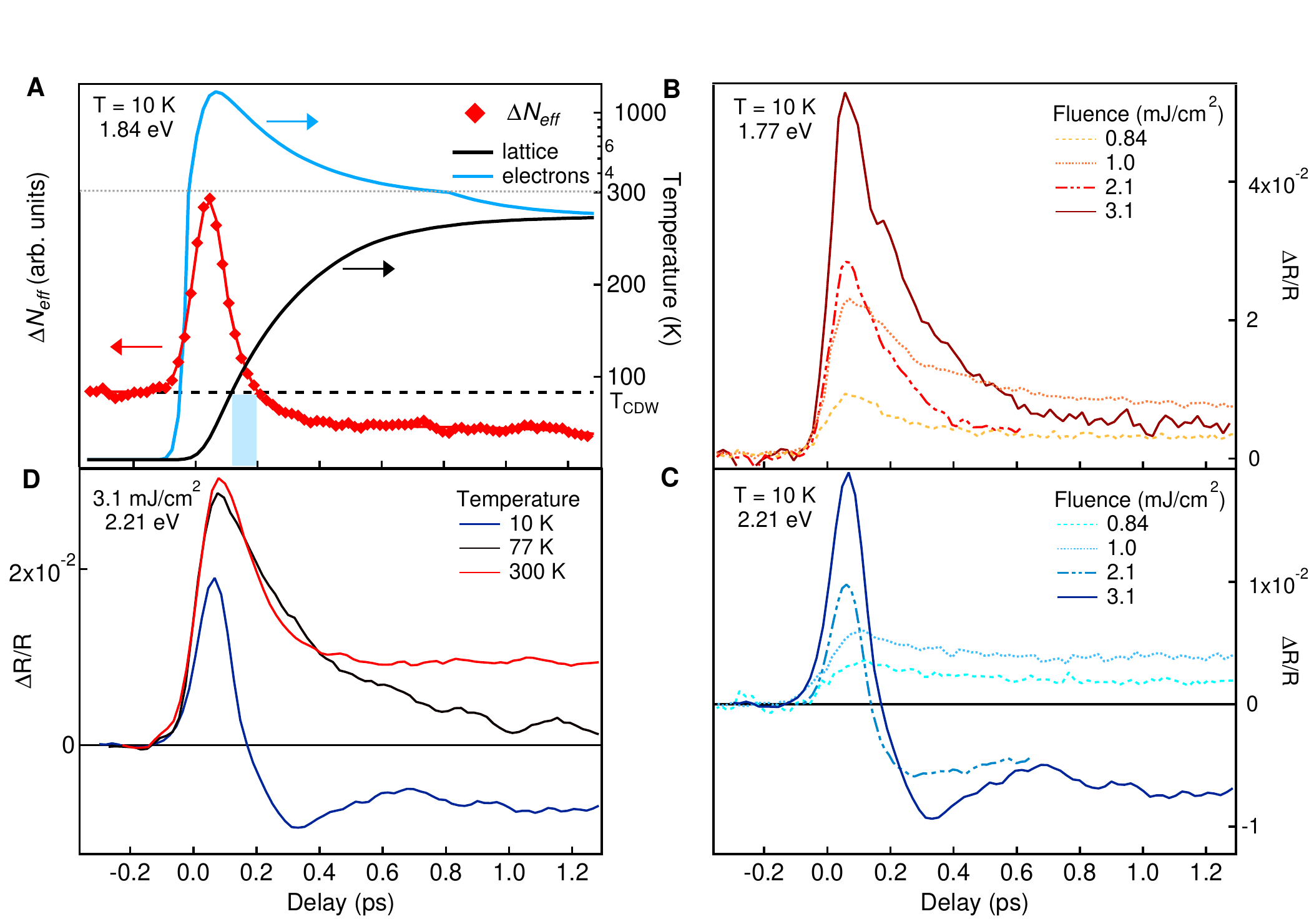}}
\caption{(\textit{A}) Time-dependent lattice (black) and electronic (blue) temperatures, obtained by solving the three-temperature model for T = 10 K, F = 3.1 mJ/cm$^2$ and an energy of 1.84 eV, and corresponding transient $N_{\textrm eff}$. The dashed line represents the zero baseline for $N_{eff}$ and the CDW transition temperature. (\textit{B-C}) Fluence dependence of the transient reflectivity traces at 10 K and 1.77 eV (\textit{B}) and 2.21 eV (\textit{C}) probing wavelengths. (\textit{D}) Temperature dependence of the transient reflectivity at 3.1 mJ/cm$^2$ pumping fluence and 2.21 eV probing wavelength.}
\label{e_ph}
\end{figure}

\end{widetext}

 To verify that indeed the behavior of the optical constants in the 2-3 eV region is mainly sensitive to the CDW melting, we measured the transient reflectivity spectra at different fluences (0.8 to 3.1 mJ/cm$^2$) and temperatures (10 K, 80 K and 300 K). In Fig.~\ref{e_ph}\textit{B} and \textit{C}, the transient reflectivity at 1.77 and 2.21 eV are shown for the different pump fluences. At 1.77 eV, the reflectivity changes are always positive, while at 2.21 eV a negative dip develops at fluences higher than 2.1 mJ/cm$^2$. In Fig.~\ref{e_ph}\textit{D}, the temperature dependence of the transient reflectivity is also shown, at a probe energy of 2.21 eV and for a fluence of 3.1 mJ/cm$^2$. Also in this case, the change of reflectivity has a negative part, associated to the CDW melting, at 10 K, while it is always positive at temperatures higher than $T_{CDW}$.
 

The time scales involved in the melting of the CDW are obtained through a transient temperature analysis using the three-temperature model~\cite{Kaganov1957, allen}, appropriate for anisotropic materials with selective electron-phonon coupling~\cite{perfetti, Mansart2010, Carbone2008}. This model describes the energy transfer from excited carriers to the lattice as an effective heat exchange between them. We obtained the temporal evolution of electronic and lattice temperatures~\cite{SI} and display them in Fig.~\ref{e_ph}\textit{A} together with the transient $\Delta N_{eff}$ at 10 K and polarization along the $c$-axis.

The CDW melting, indicated by $\Delta N_{eff}$ becoming negative at high energy, occurs after the lattice reaches a temperature above $T_{CDW}$. While the electronic temperature $T_e$ crosses $T_{CDW}$ and reaches values as high as 1100 K, the system exists in a transient state where the signature of the molten CDW state is still absent.
These simulations provide an estimate of the Eliashberg electron-phonon coupling parameter $\lambda \approx$ 0.6 and a fraction of strongly coupled modes $\approx$ 20$\%$, in agreement with the value of $\lambda$ = 0.66 determined by combined heat capacity and magnetic susceptibility measurements under pressure~\cite{Shelton1986}.
The rise time of our $\Delta R/R$ signal is in our temporal resolution range, \textit{i.e.} 40-50 fs. This, together with the three-temperature model simulations, provides an estimate of the excited electron thermalization time around 50 fs, much faster than the CDW melting time. This rules out any possible electronic melting as in this case a clear signature of the molten CDW state should be present as early as the electron thermalization time, like recently observed in TaS$_2$~\cite{cavalleri2011}.

\begin{widetext}

		\begin{figure}[ht]
		\vspace*{.05in}
\centerline{\includegraphics[width=140mm]{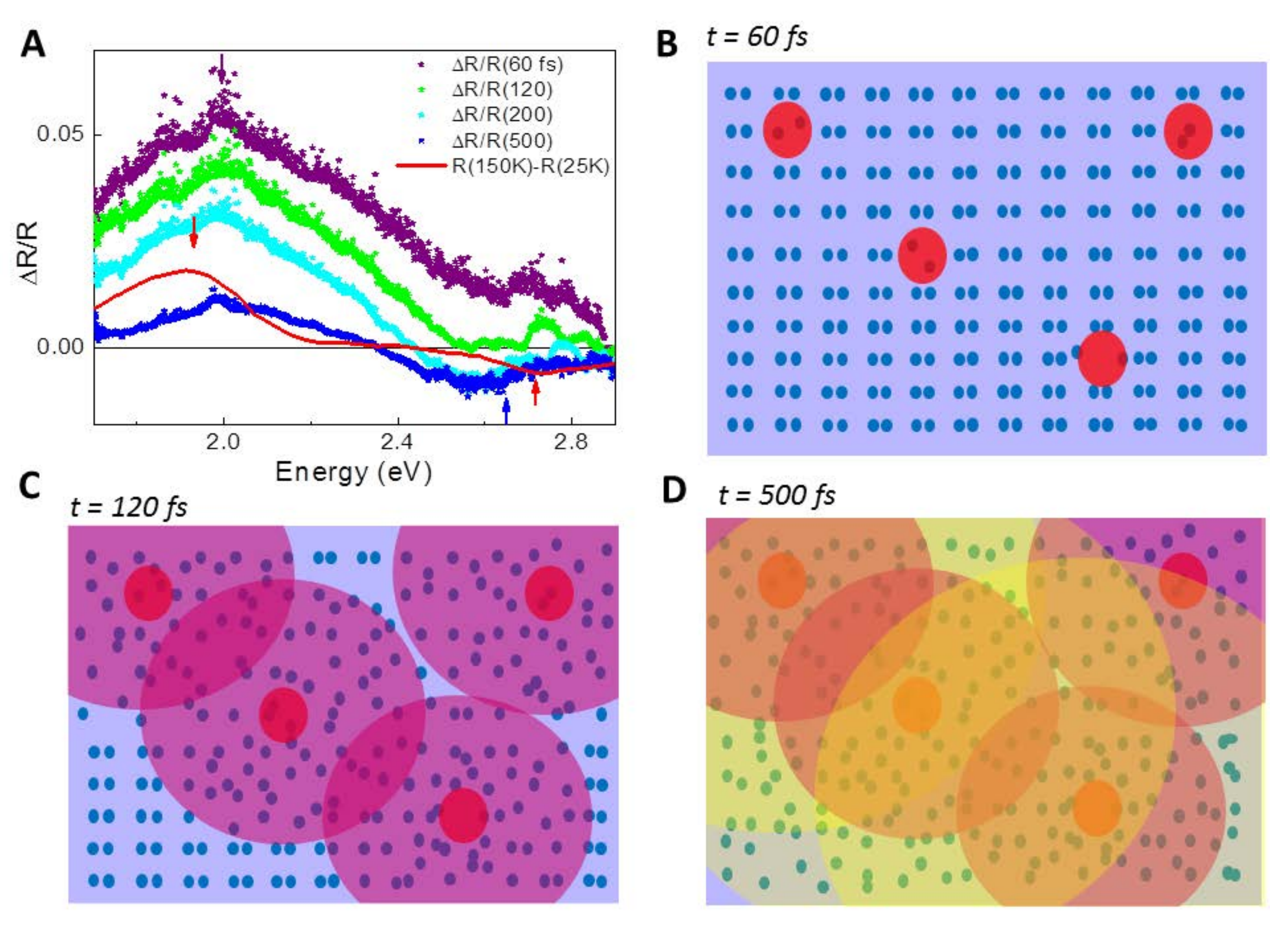}}
\caption{Transient state. (\textit{A}) Transient reflectivity spectra at 60, 120, 200 and 500 fs respectively, colored asterisks. Static temperature difference spectra, R(150 K) - R(25 K), continuous line. (\textit{B-D}) Cartoon of the CDW melting dynamics. The colored areas indicate the regions where excitation perturbs the charge order.}
\label{dessin}
\end{figure}

\end{widetext}

 To speculate on the microscopic nature of the intermediate state, we consider that during the first 200 fs the electronic structure is relaxing after being put abruptly out-of-equilibrium by the pump pulse. By comparing the spectra of the photoinduced reflectivity changes at different time delays to the difference spectra between the static reflectivity recorded at different temperatures, the impact of the out-of-equilibrium electronic structure can be estimated. In Fig. 4\textit{A}, the transient reflectivity spectra at 60, 120, 200 and 500 fs after excitation are displayed together with R(150 K) - R(25 K). When the electronic structure and the lattice are at equilibrium, the temperature difference spectrum exhibits a positive peak around 2 eV and a minimum around 2.7 eV across the phase transition (red line). This behavior is quite similar to that of the transient reflectivity after 500 fs, at which time the electronic structure is likely at equilibrium with the lattice. 
$\Delta R/R$ shows also a positive peak around 2 eV and a negative minimun around 2.7 eV; this similarity indicates that photoexcitation and temperature induce the same carriers redistribution in the solid. At early times, while the shape of the transient spectra is still similar to the static one showing a positive peak around 2 eV and a minimum around 2.7 eV, a quantitative discrepancy is more pronounced and could originate from the very large electronic temperature jump induced by laser excitation ($>$ 1000 K) compared to the static case in which the system is heated by a hundred degrees. These spectra suggest that, consistently with the three-temperature model analysis, a thermal electronic distribution of carriers is established within few tens of fs.

It is important to note at this point that while a negative reflectivity change can be associated to the phase transition, the complete melting of the CDW state can only be claimed when the optical SW, which is the cumulative effect of the positive and negative changes, becomes negative. At a fluence of 3.1 mJ/cm$^2$, around 1 carrier per unit cell is photoexcited. Photoexcited carriers create "hot areas" in which the CDW can be perturbed. This electronic perturbation diffuses in space via its interaction with the lattice and after some time, a thermally CDW molten state is reached. The delay of this process is governed by the microscopic details of the coupling between carriers and phonons in the CDW state. This scenario is pictorially represented in the cartoon in Fig. 4\textit{B-D}.

\begin{widetext}

		\begin{figure}[ht]
		\vspace*{.05in}
\centerline{\includegraphics[width=120mm]{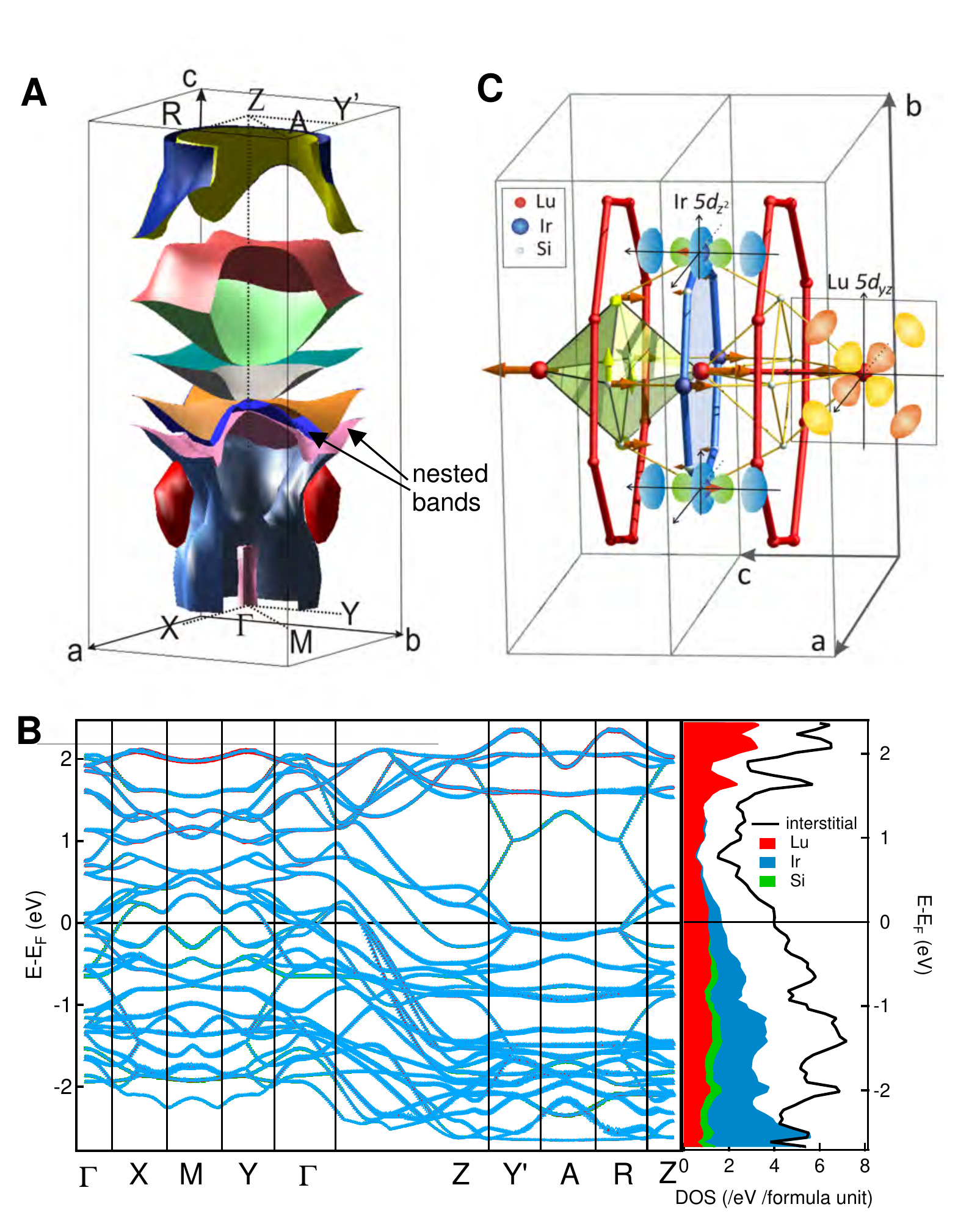}}
\caption{Electronic structure calculations. (\textit{A}) 3D map of the Fermi surface. (\textit{B}) Band structure with atomic character, and partial density of states. (\textit{C}) Representation of the main atomic orbitals (Ir $5d_{z^2}$ and Lu $5d_{yz}$) forming one of the nested bands, and atomic movement of two of the most coupled phonon modes (thick orange and yellow arrows). Lu1 atoms form chains along the $c$-axis, Lu2 the ring in the ($a,b$) plane and Lu3 (not represented) are more three-dimensionally distributed.}
\label{lda}
\end{figure}

\end{widetext}

The presented situation is very distinct from an electronically driven melting of the charge order, since in that case the negative optical $SW$ should be observable without any delay; therefore, even though an instantaneous ($<$ 60 fs) CDW melting would locally happen upon light excitation, it should propagate and melt the whole charge order within the electronic thermalization time, which we found to be around 60 fs. Such a situation is observed for example in TaS$_2$~\cite{cavalleri2011}, where a non-Peierls phase transition is claimed to take place.


Further microscopic details are obtained through {\it ab initio} electronic structure calculations~\cite{SI}, performed via both an all-electron full-potential linearized augmented plane-wave method~\cite{elk} and via pseudopotentials~\cite{qe}. Eleven bands having mixed character (color coded in Fig.~\ref{lda}\textit{B}) cross the Fermi level. The density of states (DOS) (Fig.~\ref{lda}\textit{B}) shows that the Fermi surface has mostly Ir $5d$ character. Some Lu character is found, coming from the 5$d$ orbitals strongly hybridized with both Si and Ir, producing delocalized electrons as observed in transition metal-silicides~\cite{CarboneXAS}. The Fermi surface is presented in Fig.~\ref{lda}\textit{A}, and from calculations of the generalized susceptibility, we identified two bands having nesting vectors of 2/7 and 3/7 of $c^*$, as observed by X-ray diffraction below $T_{CDW}$~\cite{Becker1999}. This suggests the presence of multiple CDWs occurring at the same transition temperature. These two nested bands account for approximately 30$\%$ of the Fermi energy DOS, in agreement with estimates based on magnetic susceptibility measurements~\cite{Shelton1986}. While one of them comes mostly from Ir $5d_{z^2}$ and Lu1 $5d_{yz}$ (Lu1 atoms being along the chains~\cite{Becker1999}, Fig.~\ref{lda}\textit{C}), the second one is mainly formed by Lu2 $5d_{yz}$-$5d_{z^2}$ and all the $5d$ orbitals of both Lu3 and Ir. This complex and three-dimensional nature of the orbitals reveals the possibility of a 3D Peierls transition. We also show in Fig.~\ref{lda}\textit{C} the motion of two stronglycoupled optical phonons, yellow and orange arrows respectively, having frequencies (1.4 THz and 2.8 THz) close to the coherent oscillations observed in the same material~\cite{Tomeljak2006}.
{\it Ab initio} calculations also estimate the electron-phonon coupling parameter of each individual lattice mode, yielding a total $\lambda \approx$ 0.6 for the 18 $\%$ of the most coupled modes, in agreement with the three-temperature model simulations.¨


The orbital occupation-number change induced by the CDW melting is signalled by the optical $SW$ transfer around a pivot energy close to 1 eV, both in static and time-resolved data.
Upon excitation, acting as a photodoping process, carriers are transferred from the Lu and Ir $5d$ states forming the CDW~\cite{SI}. This transfer depopulates the nested bands and redistributes the excited carriers above the Fermi level, increasing the low-energy $SW$ as well as the effective electronic temperature.
The observation of a delay between photoexcitation and the melting of the CDW strongly suggests the Peierls nature of the phase transition in these compounds, rather than a pure electronic effects as in different systems \cite{vo2}. In fact, while CDW are known to couple to specific lattice distortions, the intimate microscopic origin of the mechanism described in \cite{peierls} is that a periodic modulation of the lattice induces the Fermi surface instability causing the consequent charge ordering. In other words, to understand correctly the origin of such a phase transition, a notion of causality must be invoked, which is directly obtainable via time-resolved experiments. Indeed, one must wait until the atoms have moved sufficiently to suppress the distortion promoting the CDW state~\cite{Schmitt2008, Cavalleri2004}; the latency time can be of the order of one quarter of the corresponding phonon period. In our case, 200 fs gives a frequency of 1.4 THz, corresponding to the mode represented in orange arrows in Fig.~\ref{lda}\textit{C}. While the direct involvement of the lattice in the CDW formation reveals the Peierls nature of the metal-insulator transition in this material, its weakly one-dimensional electronic structure, together with the signatures of a first order CDW transition and the suggested presence of multiple CDWs, are at odds with the conventional scenario describing Peierls transitions~\cite{gruner}. Recently, the notion that Peierls transitions can happen in higher dimensions has been challenged by new ultrafast experiments in TaS$_2$ \cite{cavalleri2011}, which has a 2D charge ordering transition. Also in this case, the delay between excitation and melting has been used as the key-argument for determining the microscopic nature of the phase transition. Our experiments contribute to this exciting and current debate by showing evidence for a Peierls transition in a three-dimensional system.

The proper description of phase transitions is important for fundamental physics; in fact, phase transitions in complex systems often evolve through several unconventional intermediate states~\cite{baum}.
Critical phenomena involving abrupt perturbation of ordered systems are observed in giant land-slides~\cite{vajont} and sandpile automata~\cite{tadic}. In these cases and in CDWs, an external perturbation may induce a transient state in which the order is preserved while undergoing a collective reorganization~\cite{Middleton1992, Narayan1994}.
The speed of rock-slides, CDW slides, and flux jumps in the vortex state of type II superconductors, are self-organized critical phenomena and obey similar critical laws~\cite{Middleton1992, tinkham, Bak1987, richard}. In our experiments, the disappearance of the charge order manifests itself after a significant delay from the abrupt perturbation, and in the intermediate state the order seems preserved despite the large energy deposited.

\section{Acknowledgements}
The authors acknowledge useful discussions with D. van der Marel, J. Demsar, C. Giannetti and A. Pasquarello. The Lu$_5$Ir$_4$Si$_{10}$ single crystal have been provided by J. A. Mydosh. We thank U. Roethlisberger (EPFL, Lausanne) and ACRC (Bristol) for computer time. This work was supported by the ERC grant N$^{\circ}$ 258697 USED and by the Swiss NSF via contract $20020-127231/1$.

\title{Supporting Information: Evidence for a Peierls phase-transition in a three-dimensional multiple charge-density waves solid.}

\author{B. Mansart}
\affiliation{Laboratory for Ultrafast Electrons, ICMP, Ecole Polytechnique F\'{e}d\'{e}rale de Lausanne, CH-1015 Lausanne, CH}
\affiliation{Laboratory of Ultrafast Spectroscopy, ISIC, Ecole Polytechnique F\'{e}d\'{e}rale de Lausanne, CH-1015 Lausanne, CH}
\altaffiliation[]{barbara.mansart@epfl.ch}
\author{M. Cottet}
\affiliation{Laboratory for Ultrafast Electrons, ICMP, Ecole Polytechnique F\'{e}d\'{e}rale de Lausanne, CH-1015 Lausanne, CH}
\author{T. J. Penfold}
\affiliation{Laboratory of Ultrafast Spectroscopy, ISIC, Ecole Polytechnique F\'{e}d\'{e}rale de Lausanne, CH-1015 Lausanne, CH}
\affiliation{Laboratory of Computational Chemistry and Biochemistry, ISIC, Ecole Polytechnique F\'{e}d\'{e}rale de Lausanne, CH-1015 Lausanne, CH}
\affiliation{SwissFEL, PSI, CH-5232 Villigen, CH}
\author{S. B. Dugdale}
\affiliation{H. H. Wills Physics Laboratory, University of Bristol, Tyndall Avenue, Bristol BS8 1TL, UK}
\author{R. Tediosi}
\affiliation{D\'epartement de Physique de la Mati\`ere Condens\'ee, Universit\'e de Gen\`eve, CH-1211 Gen\`eve 4, CH}
\author{M. Chergui}
\affiliation{Laboratory of Ultrafast Spectroscopy, ISIC, Ecole Polytechnique F\'{e}d\'{e}rale de Lausanne, CH-1015 Lausanne, CH}
\author{F. Carbone}
\affiliation{Laboratory for Ultrafast Electrons, ICMP, Ecole Polytechnique F\'{e}d\'{e}rale de Lausanne, CH-1015 Lausanne, CH}

\maketitle

\section{Experimental setup}
\label{exp}

We performed an experimental study of the transient optical properties of Lu$_5$Ir$_4$Si$_{10}$ using pump-probe broadband reflectivity. Our experimental setup, located at the Laboratory of Ultrafast Spectroscopy, EPFL, Switzerland, is presented in Fig. S1. An amplified Ti:sapphire laser system provides 1.55 eV pulses of 40 fs duration at a repetition rate of 1kHz. This laser beam is separated into two paths by a beamsplitter. The pump beam, whose energy is fixed at 1.55 eV, and the probe beam which is converted into a supercontinuum by a CaF$_2$ nonlinear crystal. The probe spectrum covers a range between 1.5 and 3.2 eV. The fluence of the white light probe is several orders of magnitude lower than the pump beam and has no effect on the measurements. The pump pulse is used to excite the system at the time $t=0$, and the consequent variations of reflectivity are recorded with a delay $t$ by changing the optical path difference between pump and probe (which can be achieved by setting a variable delay in the probe arm).

Small reflectivity changes (down to the order of 10$^{-3}$) are detectable owing to the signal-to-noise ratio of 10$^4$, obtained by the use of an optical chopper on the pump beam which reduces its repetition rate by a factor of two. This allows the recording, for each time delay, of a single-shot spectrum with and without the pumping pulse. Furthermore, the variations of the white light intensity are taken into account using a reference signal, measured just before the sample position.

Two identical spectrometers are used to detect the transient reflectivity signal and the reference one. By dispersing the incident light with optical gratings and sending it to an array detector, we obtain the complete spectrum for each single pulse with an energy resolution better than 1 meV. Subsequently, the group velocity dispersion introduced by the chromatic aberration of the optics is corrected before the transient reflectivity analysis.

The pump-beam diameter on the sample surface was 250 $\mu$m, while that of the probe was set to 100 $\mu$m in order to probe a uniformly excited area. Both spot sizes and spatial overlap were rigorously verified using a beam profiler and a charge-coupled device camera. During this experiment, the pump fluence was varied between 0.8 and 3.1 mJ/cm$^2$.

Prior to these experiments, the sample was oriented using X-ray diffraction and polished to obtain good optical quality. Its surface contained both (001) and (100) directions. Therefore, pump and probe polarizations were changed independently with quarter-waveplates and were oriented either parallel or orthogonal to the $c$-axis (containing the Lu chains). Finally, in order to cool down to 10 K and precisely monitor the sample temperature, the latter was placed in a liquid He cryostat.

Such kind of experimental setup has been successfully employed to investigate photoexcited states in strongly correlated systems, such as the charge-transfer insulator CuGeO$_3$~[S1], as well as in bulk graphite and graphene paper~[S2].

\section{Spectral Weight analysis}
\label{SW1}

The definition of the optical spectral weight ($SW$) in a solid, given in Eq. (1) of the main article, makes clear that the evaluation of $N_{eff}$ requires the knowledge of the real part of the optical conductivity down to zero frequency. Since this is not experimentally feasible, common procedures (including extrapolations or modeling based on the Drude-Lorentz formalism) are invoked to obtain the whole $\sigma_1(\omega)$ spectrum. Recently, it has been shown mathematically that according to the theorems on analytical continuation of holomorphic functions, and taking advantage of the Kramers-Kronig relation between the real and the imaginary part of the complex optical conductivity ($\sigma(\omega) = \sigma_1(\omega)+i\sigma_2(\omega)$), the partial $SW$ integral can be re-written as a function of a limited spectral interval in the form of Carleman-Goluzin-Krylov equations~[S3]:

\begin{equation}
SW(\Omega_c) = Re~\{\lim_{n\rightarrow\infty}\int^{\omega_{max}}_{\omega_{min}} u_n(\omega^\prime,\Omega_c)\sigma(\omega^\prime)d\omega^\prime\}
\label{Neff2}
\end{equation}

where $u_n(\omega^{\prime},\Omega_c) = \int_{0}^{\Omega_c}Q_n(\omega^{\prime},\omega)d\omega$ are properly chosen kernel functions. In the case of real experimental data (with finite energy resolution and statistical noise), Eq. (\ref{Neff2}) cannot be strictly applied, and a numerical procedure based on the idea of the analytical continuation has to be invoked in order to obtain the $SW$ integral in a model-independent fashion~[S4]. A code called ``Devin'' has been developed which allows the $SW$ integral to be extracted from the complex dielectric function in a limited data range in a model-independent fashion~[S4].

 We performed a Drude-Lorentz analysis of our experimental data, and verified the results via the model-independent routine ``Devin''. The approach is described below in more detail.

 To obtain the complex dielectric function, we fitted our transient reflectivity data to a highly flexible Drude-Lorentz function~[S5]; the underlying Drude-Lorentz model of the optical data is based on the knowledge of the static optical spectra between a few meV and 4.6 eV, data taken from Ref.~[S6].

For the fitting of the transient data, the Lorentz oscillators in our experimental range are allowed to change in order to reproduce the dynamical reflectivity. This procedure yields the transient $\epsilon_1$ and $\epsilon_2$, displayed together with the static spectra in Fig. S2.
In our experimental frequency range, the real and imaginary part of the dielectric constant have rather similar absolute values (Fig. S2\textit{A} and \textit{E}). For this reason, the optical reflectivity, defined as $R = \mid\frac{1-\surd\epsilon}{1+\surd\epsilon}\mid^2$, is equally sensitive to the reactive and the absorptive components of the dielectric function. As a consequence, the complex $\epsilon$ obtained through the Drude-Lorentz fitting of the optical reflectivity matches very well the one obtained via spectroscopic ellipsometry (see Fig. S2\textit{A} and \textit{E} and Ref.~[S6]).
The transient dielectric function is displayed in Fig. S2\textit{(B-D, F-H}) for light polarized along the $a$-axis and the $c$-axis at 10 K, and for light polarized along the $c$-axis at room temperature.

Such a Drude-Lorentz modeling of the data also yields the spectral weight; the overall static $N_{eff}$ spectrum for both light polarizations is displayed in Fig. S3. In the same plot, we show the derivative of $N_{eff}$ as a function of the number of carriers which exhibits an inflection point between 1 and 1.5 eV for light polarized along the $c$-axis; this inflection point is found to be between 10 and 15 carriers. Since Lu has a $5d^1$ electronic configuration, and we found a strong hybridization with at least one or two Ir $5d$ bands, we expect between 10 and 20 electrons to form the material conduction band and therefore the low energy $c$-axis conductivity should be mainly sensitive to these orbitals.

By integrating the transient conductivity, we calculated $N_{eff}$ at each time step and obtained the time evolution of $N_{eff}$ via Drude-Lorentz model, presented as 3D color maps in Fig. 2(\textit{A-C}) of the main article.

In order to verify that the transient changes of the $SW$ we obtained by extrapolating the Drude-Lorentz model down to low frequency are robust and do not depend on the latter extrapolation, we fed into the ``Devin'' routine the real and imaginary  parts of the dielectric function as an input in order to generate the $SW$ up to a given frequency cut-off. Through this procedure it is also possible to obtain an estimate of the uncertainty related to the knowledge of the spectrum being in a limited range with a limited resolution and with a given noise level. Because we are interested in the
relative changes of the $SW$ induced by pump pulses, we use as an error our estimated accuracy in the determination of the transient dielectric function, which is of the order of $10^{-4}$, while the resolution of our spectrometer is approximately 1 meV. The transient dielectric function is fed into ``Devin'' and the model-independent $SW$ changes are shown in Fig. 2(\textit{A-B}) of the main article as white lines with errorbars. The fact that both these analyses give consistent results validate our description of the transient optical reflectivity.

The reason why we can successfully obtain information on the CDW-induced spectral weight transfer from a limited spectrum in the visible light region is that, in this sample, the optical anisotropy, related to the Lu one-dimensional chains, strongly manifests itself around 2 eV. This is visible in the static reflectivity spectrum (Fig. 1 of the main article), and also in the temperature-dependent dielectric constants obtained via ellipsometry and reported in Ref.~[S6], which in fact show a particularly large CDW-induced change between 1.5 and 2.5 eV.

Also, in strongly correlated solids, a phase transition can induce spectral weight transfers across a very broad energy range, extending to the visible light region. In these cases, it has been shown that the spectral region in the proximity of the screened plasma frequency of the material is very sensitive to these shifts~[S7, S8]. Here, the screened plasma frequency along the $c$-axis (direction of the charge-density-wave) lies within our experimental range (zero crossing of $\epsilon_1$ in Fig. S3\textit{A}).

\section{Three-temperature model}
\label{3TM}

The two-temperature model was first introduced by Kaganov \textit{et al.}~[S9] in order to describe the energy transfer between the electron and lattice subsystems, in the case where the electrons are driven to a higher temperature than the phonons. Its validity relies on the electronic temperature being of the order of the Debye temperature (366 K in Lu$_5$Ir$_4$Si$_{10}$~[S10]) and on the isotropy of the electron-phonon (e-ph) coupling function. For anisotropic materials like cuprates~[S11] or iron-pnictides~[S12], a selective coupling between electrons and a subset of the total phonon modes may be taken into account using the three-temperature model (3TM), governed by the following equations:

\begin{equation}
\begin{array}{l}
\label{eq_3TM}
2 C_e \frac{\partial T_e}{\partial t}= \frac{2(1-R)}{l_s}I(t)+\frac{\partial}{\partial z}(\kappa_e \frac{\partial T_e}{\partial z})-g(T_e-T_h)\\
~\\
\alpha C_L \frac{\partial T_h}{\partial t} = g(T_e-T_h)-g_c(T_h-T_c)\\
~\\
(1-\alpha)C_{L} \frac{\partial T_c}{\partial t}= g_c(T_h-T_c)\\
~\\
\end{array}
\end{equation}

where $T_e$, $T_h$ and $T_c$ are the temperatures of electrons, efficiently coupled phonons (``hot phonons") and remaining modes (``cold phonons"), respectively. $C_e$=$\gamma T_e$ is the electronic specific heat, $C_L$ is that of the lattice. $\alpha$ is the fraction of efficiently coupled modes, $R$ is the static reflectivity and $l_s$ is the penetration depth (both at the pump energy). $I(t)$ the pumping intensity, and $\kappa_e$ is the electronic thermal conductivity (which is ignored in our case). Indeed, the electronic diffusion term $\frac{\partial}{\partial z}(\kappa_e \frac{\partial T_e}{\partial z})$
changes neither the relaxation rates nor the e-ph thermalization time. The constant $g$ governs the energy transfer rate from electrons to hot phonons, and is related to the second moment of the Eliashberg
function $\lambda \left\langle \omega^2 \right\rangle$ through $g=\frac{6\hbar\gamma}{\pi k_B} \lambda \left\langle \omega^2 \right\rangle$~[S13], $\lambda$ being the dimensionless e-ph coupling constant.
$g_c$ governs the energy relaxation from hot phonons to the remaining lattice.

To perform the 3TM simulations, we used an iteration procedure, calculating at each time step the depth-dependent temperature profiles. At each depth and time step, we iterate the electronic and lattice part of the specific heat. The electronic part is taken as $C_e=\gamma T_e$, with $\gamma$=23.42~$mJ\cdot mol^{-1}\cdot K^{-2}$~[S10], and the lattice part is taken from Ref.~[S14].

The e-ph coupling constant $\lambda$ was obtained via these simulations, given a logarithmic average of the phonon energies $\left\langle \omega^2 \right\rangle$ of 30 meV, obtained by static optical spectroscopy in Ref.~[S6]. The e-ph coupling constant should in any case depend on the probing energy; as a check, we simulated the experimental data for several probe wavelengths in order to show the result constancy in the experimental energy range. The constant obtained through these simulations is given in Fig. S4\textit{C}, from which we may notice its energy-independence in the range 1.6 --2 eV. 

 In fact, as this energy range is far from the electronic inter-band transitions (except very close to 2 eV, see the Lorentz oscillator positions in Fig. 1\textit{B} of the article), the probe is mainly sensitive to intra-band transitions. 
 
 Considering $N_{eff}$, both experimental and theoretical determinations show that the carriers involved in the optical conductivity at 1.5 eV are mainly the Lu 5$d$ and some Ir 5$d$, which mainly belong to the conduction band of the material; in fact, the plasma edge of the solid is close to 1.7 eV along the $c$-axis. Therefore, even though the numerous possible inter-band transitions should give a non-negligible contribution to the optical absorption, screening effects prevent the existence of long lived excited states at those energies (1.5 eV) in such a metallic system, and the typical electron-hole lifetime should not exceed 10-20 fs. Therefore, on the time-scales of our experiments, $>$ 50 fs, most if not all of the light excitation creates electron-hole pairs which relax instantaneously back in the conduction band with an excess energy that can be described by an effective temperature. Consequently, applying the 3TM in this energy range allows the measurement of the coupling between electrons at the Fermi level and lattice.

The linear dependence of the fast relaxation time as a function of the maximum electronic temperature reached (see Fig. S4\textit{A}), as expected considering this relaxation as due to e-ph coupling~[S13], reflects also the applicability of the 3TM in our case.

We may notice the excellent agreement between the parameters obtained from this effective temperature model and the calculated partial e-ph coupling constants presented in the following section. Indeed, both the fraction of efficiently coupled phonons (Fig. S4\textit{D}) and the sum of their partial e-ph coupling (Fig. S4\textit{C}) may be retrieved by 3TM simulations, enhancing our confidence in the reliability of this model in describing our transient data.

 

\section{Electronic structure calculations}
\label{LDA}

Two separate sets of calculations were performed, the first using the highly accurate all-electron full-potential linearized augmented plane-wave method (often referred to as FP-LAPW), as implemented in the ELK code~[S15], and the second using pseudopotentials, as implemented within the QUANTUM ESPRESSO code~[S16], for the calculation of the phonons and e-ph coupling.

The lattice constants used were the experimental ones ($a=12.4936$\AA, $c=4.1852$\AA), as were the atomic positions~[S17], giving a unit cell of two formula units (38 atoms). The calculations made using the ELK code [S15] had a cutoff (for planewaves in the interstitial region) determined by $k_{\rm max}=7.0/R_{\rm min}$, where $R_{\rm min}$ is the smallest muffin-tin radius. These muffin-tin radii were Lu (2.3328 a.u.), Ir (2.3654 a.u.) and Si (1.8369 a.u.). The self-consistent cycle was carried out on a mesh of 64 $k$-points in the irreducible (one sixteenth) of the Brillouin zone, and the Fermi surface
evaluated from the converged potential on a denser mesh of 1936 $k$-points within one eighth of the Brillouin zone.

The calculations presented are scalar-relativistic and omit spin-orbit coupling (after the effect of its inclusion was investigated and found to be negligibly small at the Fermi level.) The exchange-correlation functional was that of Perdew-Wang/Ceperley-Alder, Ref.~[S18].

The calculated band structure in the energy range [-8 eV, +2 eV] is given in Fig. S5\textit{A}. In this range, the latter is composed of strongly hybridized Lu $5d$ and $4f$, Ir $5d$ and Si $2p$, the Lu $4f$ giving localized density of states around -4.5 eV, while electrons at the Fermi surface come mainly from Lu $5d$, Ir $5d$ and Si $2p$ as shown in Fig. S5\textit{B} and Fig. 5 of the main article.
A large amount of ``interstitial" electronic density is also found at the Fermi level, indicating their delocalized nature as well as the strong hybridization degree in this system.

The e-ph coupling resulting from the mass enhancement may be calculated from the density of states (DOS) at the Fermi level, following the formula $N(E_F)=N_b(E_F) \frac{m}{m^*}=N_b(E_F) (1+\lambda)$, where $N(E_F)$ is the renormalized DOS and $N_b(E_F)$ the bare one, both at the Fermi level $E_F$; $m$ is the free electron mass and $m^*$ the effective one. The renormalized DOS obtained by specific heat measurement being 0.26/eV/atom/spin, (i.e. 19.76/eV), and the bare DOS calculated here of 11.02/eV if one considers that 30$\%$ of the Fermi surface is gapped, one obtains $(1+\lambda)=1.79$ giving an e-ph coupling constant $\lambda=0.79$.

We also present in Fig. S6 the real and imaginary parts of the calculated susceptibility for two bands, corresponding of the Fermi surfaces indicated by arrows in Fig. 5\textit{A} of the main article. One may notice the peak at two different wavevector, in both real and imaginary parts. These peaks appear at nesting wavevectors which are different between the two presented bands, indicating the formation of several charge-density waves.

For the phonon and e-ph coupling calculations, linear response within the pseudopotential approach provided by the QUANTUM ESPRESSO code was used [S16]. The Lu pseudopotential had the $f$ electrons
as core states, and the resulting electronic structure was verified against that produced by the all-electron ELK calculation. Both ultrasoft and norm-conserving pseudopotentials were used, the latter needed for the calculation of the dielectric function with the \verb+EPSILON+ package within QUANTUM ESPRESSO. The cut-offs for the energy and charge density were 40 Ry and 400 Ry, respectively. Phonons at the $\Gamma$ point were calculated from self-consistent calculations performed on a (6,6,16) $k$-point mesh and e-ph parameters were evaluated on a denser mesh of (12,12,32). For the calculations of dielectric function, the routine \verb+EPSILON+ within the QUANTUM ESPRESSO package was used. Here, Brillouin zone integrations were carried out on a mesh of over 6000 $k$-points with a Gaussian smearing of 0.25 eV applied to the evaluation of the inter-band contribution.

We present the partial e-ph coupling constants for each of the 114 modes of the system as a function of the mode energy at $q=0$ in Fig. S7\textit{B}. Two strongly coupled modes are found having respectively a partial e-ph coupling constant of 0.166 and 0.131, involving mainly Si atom motion since the latter are the lighest elements of the crystal structure. One of these two modes is represented with yellow arrows in Fig. 5\textit{C} of the main paper. The sum over all modes gives $\lambda=0.97$, so in good agreement with the value obtained from the DOS renormalization at the Fermi level
(when one considers that the electron-phonon calculation has only been made at the $\Gamma$ point).

The histogram of partial e-ph coupling constants is shown in Fig. S7\textit{A}. From such a representation, one is able to extract the partial e-ph coupling of some percentage of the most efficiently coupled modes, as represented with a red-dashed line in Fig. S7\textit{A}. This area contains 21 modes, (or 18 $\%$ of the phonons), and summing up their $\lambda$ constant one obtain 0.62. It indicates that the 18 $\%$ of the phonon modes having the largest e-ph coupling have, all together, an e-ph coupling constant of 0.62. This observation is in excellent agreement with the results of 3TM simulations, also indicating the reliability of these electronic structure and phonon calculations.
This value of $\lambda$ gives a superconducting transition temperature of $\approx$ 9 K using the McMillan formula~[S19], as experimentally observed in these system above the critical pressure, and suggests that conventional e-ph mediated superconductivity takes place in Lu$_{5}$Ir$_{4}$Si$_{10}$, obeying Bardeen-Cooper-Shrieffer theory. Indeed, since upon photoexcitation the CDW is molten, and therefore does not compete anymore with superconductivity (as in the system under pressure), obtaining a similar $\lambda$ value may reflect effectively the order parameter giving rise to superconductivity and not the one related to Peierls ordering.

\begin{widetext}

		\begin{figure}[ht]
		\vspace*{.05in}
\centerline{\includegraphics[width=120mm]{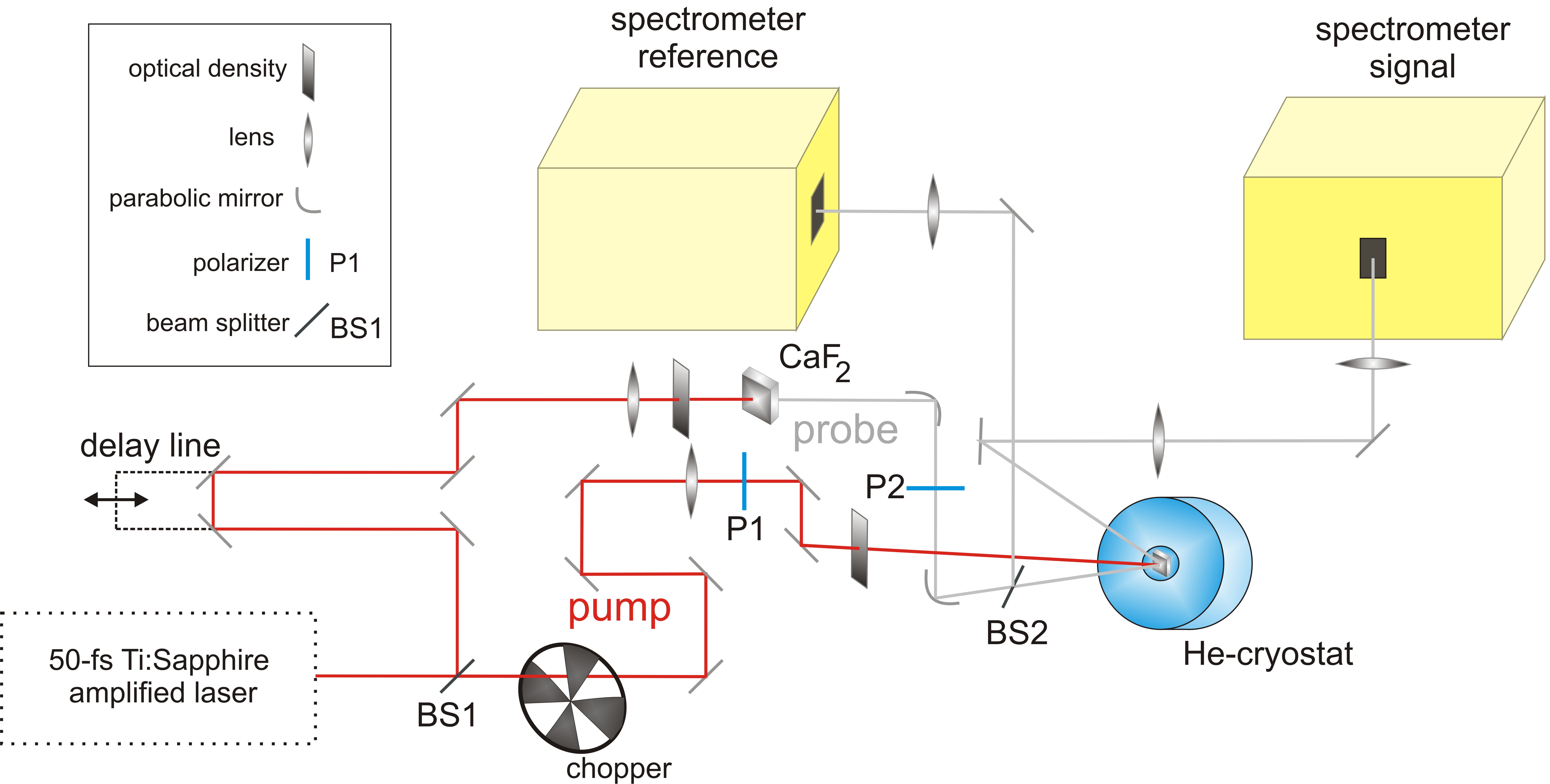}}
\caption{Experimental Setup.}
\label{reflectivity}
\end{figure}

\begin{figure}[ht]
\vspace*{.05in}
\centerline{\includegraphics[width=120mm]{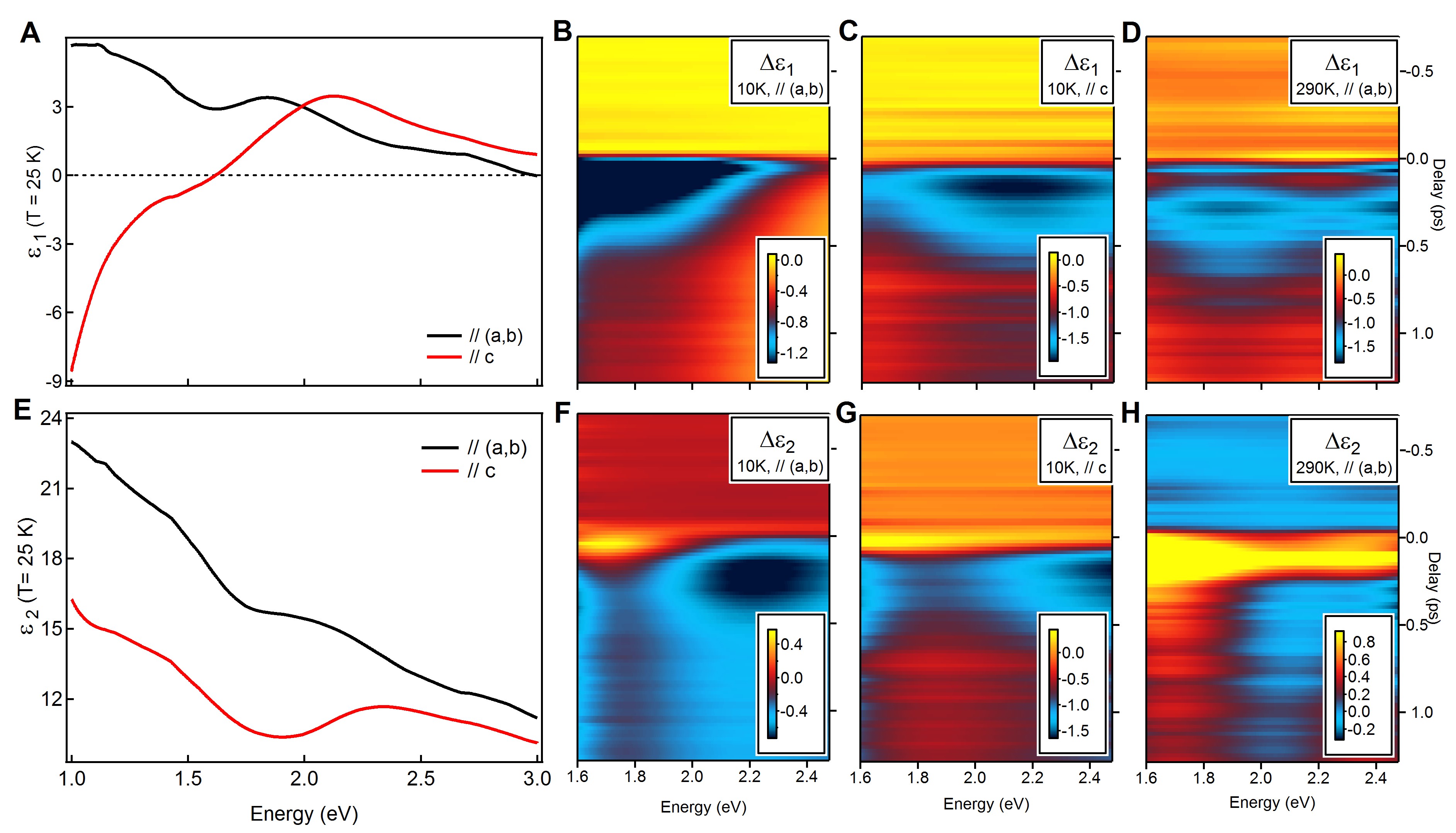}}
\caption{(\textit{A}) and (\textit{E}) Static dielectric function for both light polarizations. Color maps: transient dielectric function at different temperatures and along the different orientations.}
\label{e_ph}
\end{figure}

\begin{figure}[ht]
\vspace*{.05in}
\centerline{\includegraphics[width=150mm]{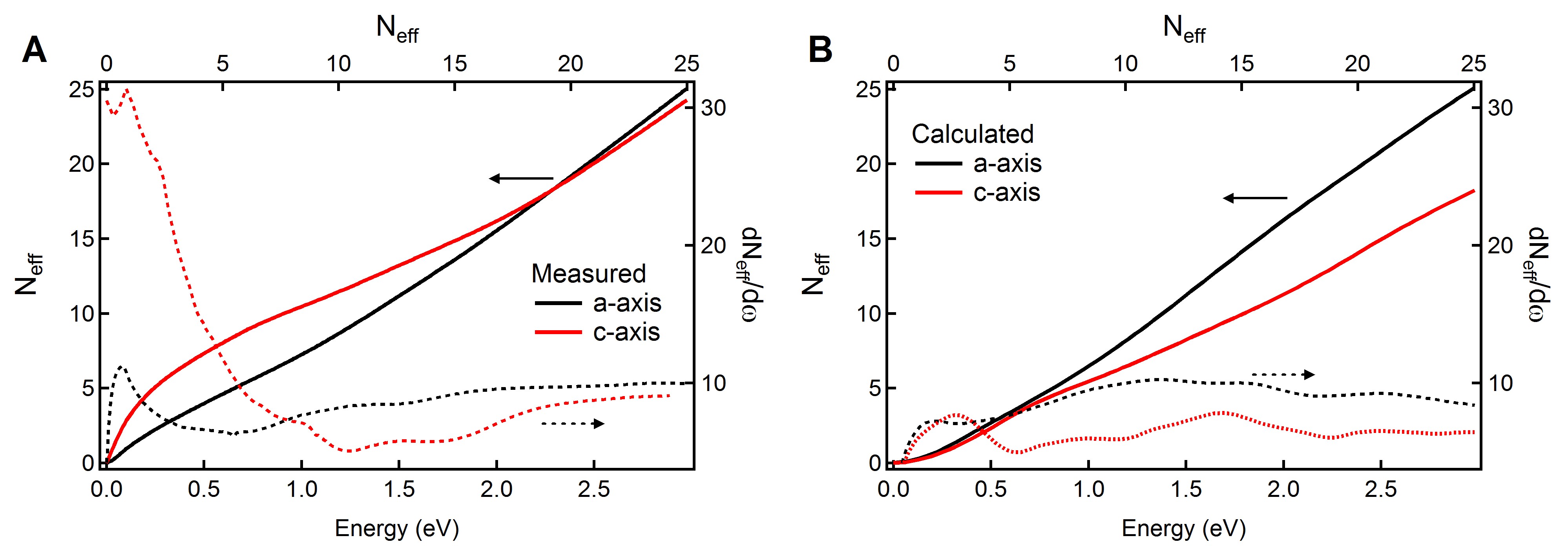}}
\caption{Effective number of carriers $N_{eff}$ as a function of energy, and its derivative $\frac{d N_{eff}}{d \omega}$ plotted as a function of $N_{eff}$; (\textit{A}) measured data, and (\textit{B}) obtained through electronic structure calculations described in the text.}
\label{SW}
\end{figure}

		\begin{figure}[ht]
		\vspace*{.05in}
\centerline{\includegraphics[width=150mm]{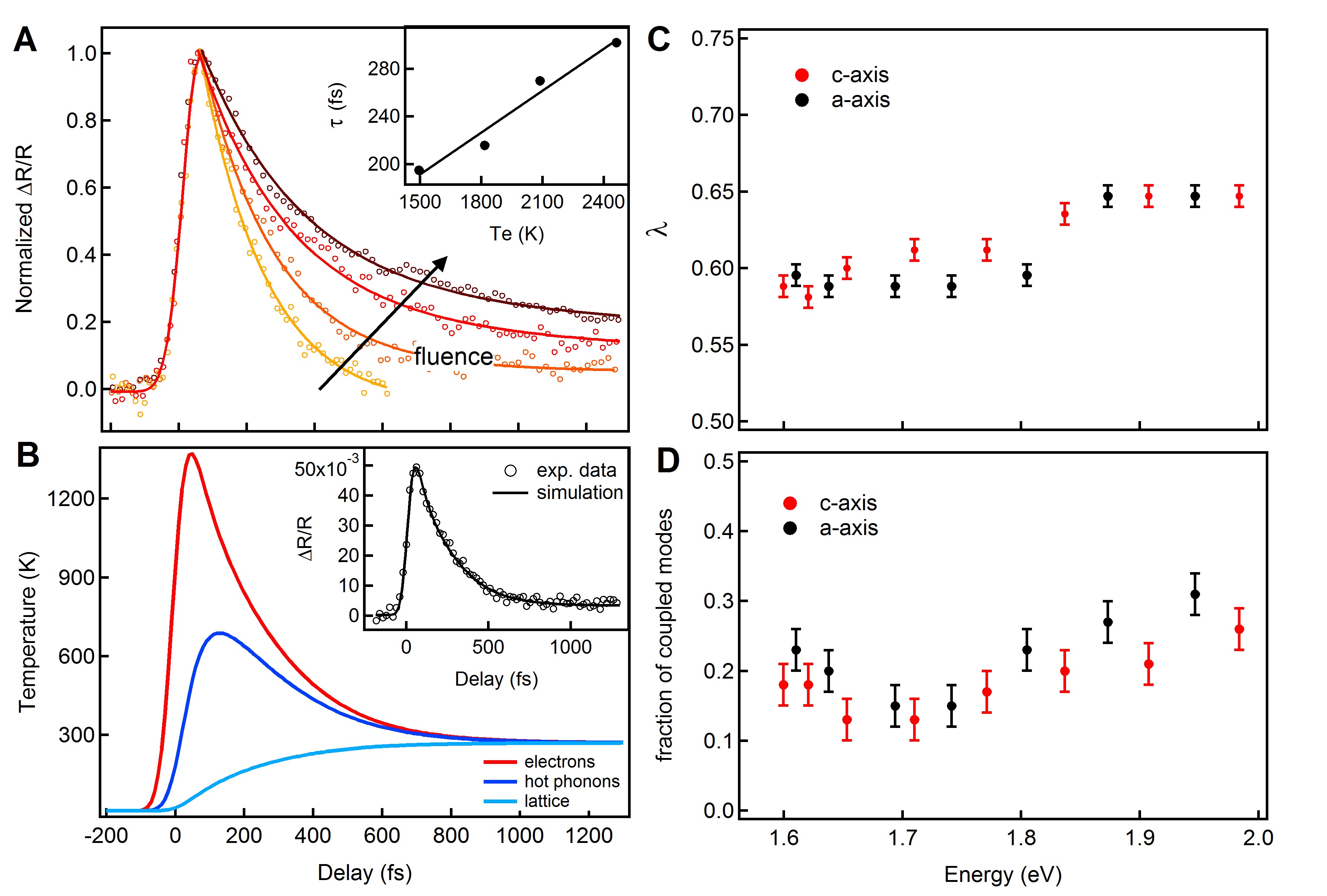}}
\caption{Three temperature model analysis. (\textit{A}) Fluence dependence of the signal obtained at 10K, with pump and probe parallel to $c$-axis, at an energy of 1.7 eV; in the inset the fast decay time, associated with the electron-phonon coupling time, is plotted as a function of the maximum electronic temperature reached in average in the skin-depth together with its linear fit. (\textit{B}) Electron, hot phonon and cold phonon temperatures as a function of time delay obtained by simulating the transient reflectivity shown in open circles in the inset, with the simulation as the solid line; this curve corresponds to a geometry where both pump and probe were parallel to the $c$-axis, at a pump fluence of 3.1 mJ/cm$^2$, and a probing energy of 1.7 eV. (\textit{C}) electron-phonon coupling constant obtained from the simulations, for probe polarization parallel to $c$-axis or within the ($a,b$) plane, simulated at a pump fluence of 3.1 mJ/cm$^2$ as a function of the probing energy; (\textit{D}) corresponding fraction of coupled modes.}
\label{lda}
\end{figure}

\begin{figure}[ht]
\vspace*{.05in}
\centerline{\includegraphics[width=150mm]{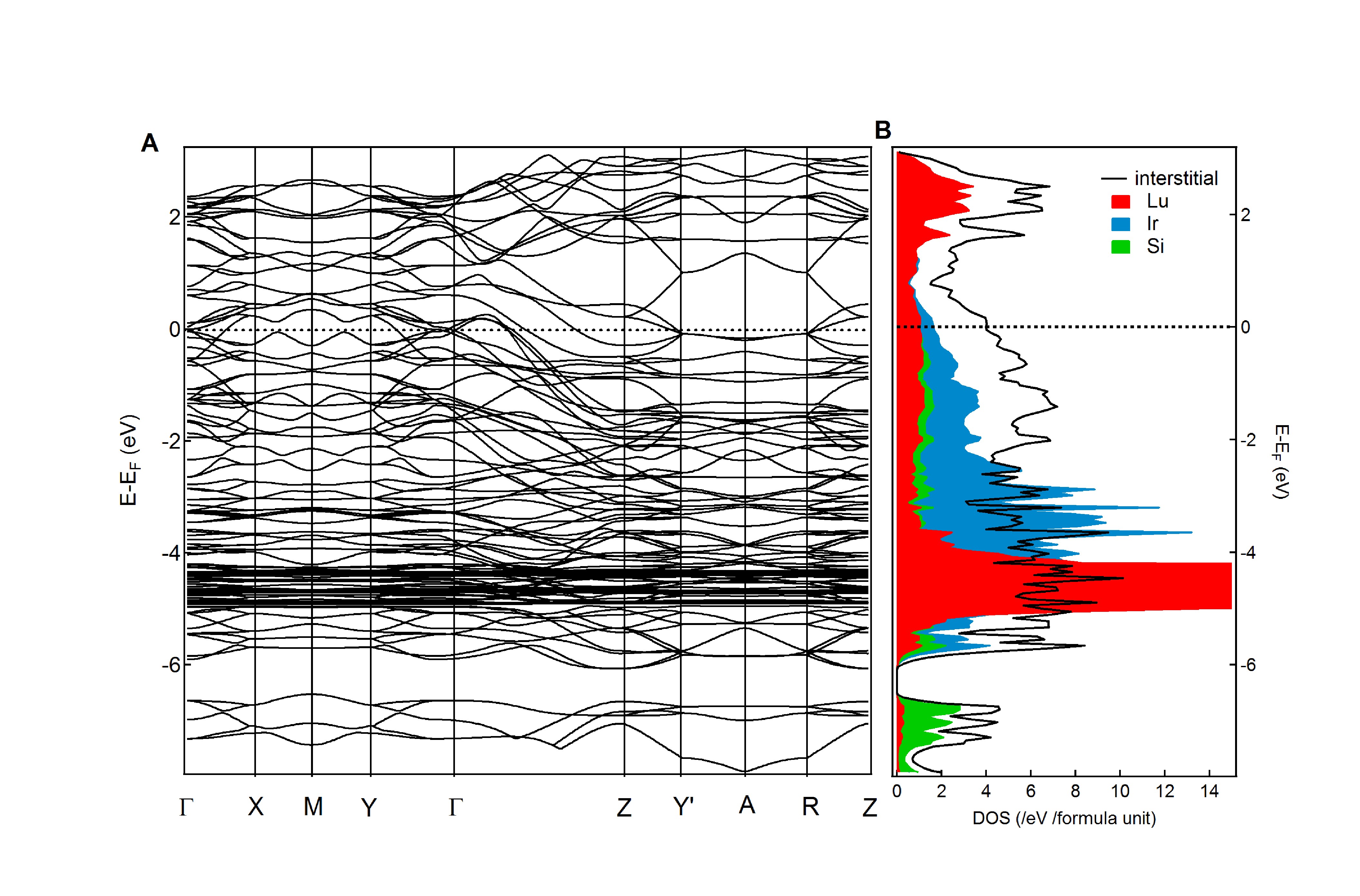}}
\caption{Electronic band structure. (\textit{A}) Calculated electronic band structure using the ELK code. (\textit{B}) Density of states (note that the contribution of the localised Lu 4$f$ states is truncated).}
\label{DOS}
\end{figure}

\begin{figure}[ht]
\vspace*{.05in}
\centerline{\includegraphics[width=80mm]{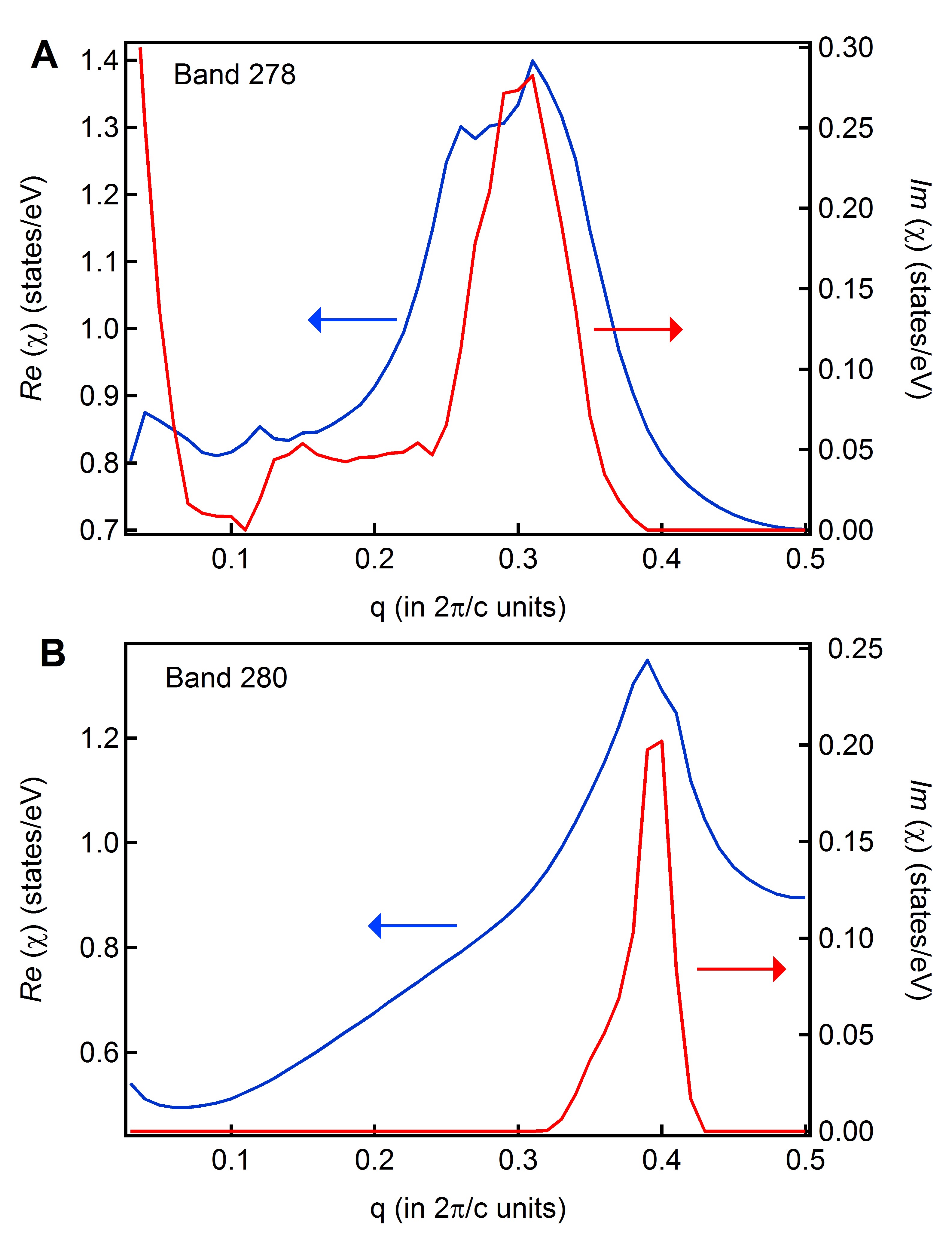}}
\caption{Calculated real and imaginary part of the susceptibility, for two of the bands close to the Fermi level (labeled 278 and 280).}
\label{Chi}
\end{figure}

\begin{figure}[ht]
\vspace*{.05in}
\centerline{\includegraphics[width=130mm]{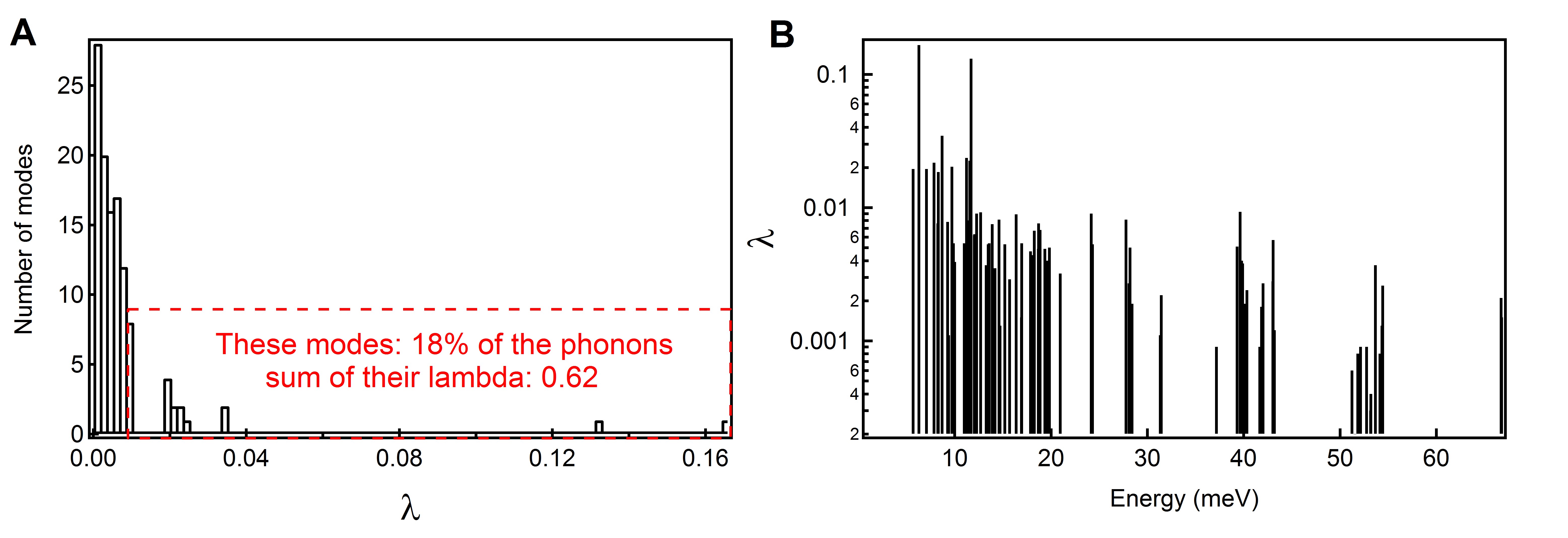}}
\caption{Calculated electron-phonon coupling. (\textit{A}) Histogram of the electron-phonon coupling constants $\lambda$; the red area represents the 18$\%$ more coupled modes, whose total lambda gives 0.62. (\textit{B}) Energy distribution of the $\lambda$.}
\label{lambda}
\end{figure}

\end{widetext}

\end{document}